\documentclass[aps,prd,floatfix,nofootinbib]{revtex4} % preprint, twocolumn, groupedaddress, nofootinbib
\usepackage{graphics}
\usepackage{graphicx}

\usepackage{latexsym}
\usepackage[cp866]{inputenc}
\usepackage[english]{babel}
\input epsf

\begin{document}

\title{\boldmath
Estimate of the $S$-wave $D^*K$ scattering length in the isospin-0
\\ channel from Belle and LHCb data}
\author{N. N. Achasov\,\footnote{achasov@math.nsc.ru}
and G. N. Shestakov\,\footnote{shestako@math.nsc.ru}}
\affiliation{\vspace{0.2cm} Laboratory of Theoretical Physics, S. L.
Sobolev Institute for Mathematics, 630090, Novosibirsk, Russia}

%\date{}

\begin{abstract}
%------------------------------------------------------------------------------------------------------------------------------------------------

It is shown that the Belle and LHCb data on the interference of the
amplitudes of the $S$- and $D$-partial waves in the decays
$D_{s1}(2536 )^+\to D^{*+}K^0_S$ and $D_{s1}(2536)^-\to\bar
D^{*0}K^-$ allow us to obtain an estimate of the $S$-wave $D^*K$
scattering length in the channel with isospin $I=0$:
$|a^{(0)}_{D^*K}|=(0.94\pm0.06)$ fm. The possibility of explaining
the found value by the contribution of the $D_{s1}(2460)$ resonance
is discussed. The decay of $B_{s1}(5830)\to B^*\bar K$ is also
briefly discussed.

%------------------------------------------------------------------------------------------------------------------------------------------------
\end{abstract}

\maketitle
%------------------------------------------------------------------------------------------------------------------------------------------------
\section{\boldmath Introduction}

Recently, much attention from theorists has been paid to
calculations of the $S$-wave scattering lengths of charmed mesons on
light pseudoscalar mesons (see for details Refs. \cite{Gu09,Ge10,
Li11,Li13,Gu18,Gu19,Hu22,Al23,Ik23,To23,Kh24,Li24} and references
herein). The results were obtained using combinations of lattice
calculations \cite{Li13,Gu19}, effective chiral theory
\cite{Gu09,Ge10,Li11,Gu18,Gu19,Hu22}, and other methods
\cite{Al23,Ik23,To23,Kh24,Li24}. Conventional experiments on the
scattering of $D$ mesons are impossible because of the short
lifetime of these particles. In principle, a method based on
measuring femtoscopic correlation functions of hadron pairs allows
one to obtain information about strong interactions of charmed
mesons and to check the results of scattering length calculations;
see in this connection Refs. \cite{Al23,Ik23,To23,Kh24,Li24,AL17,
AL22,Ka22,Ad24,AL24}. Recently, the ALICE Collaboration \cite{AL24}
obtained for the first time the data on the correlation functions of
the pairs $D^\pm\pi^\pm$, $D^{*\pm}\pi^\pm$, $D^\pm K^\pm$, and
$D^{*\pm} K^\pm$ for all charge combinations. In this experiment,
the strong interaction between charged mesons manifested itself as a
residual interaction against the background of a significant Coulomb
contribution. In Ref. \cite{To23}, it was noted that measuring the
correlation functions of the $D^{*0}K^+$ and $D^{*+}K^0$ pairs is a
more difficult task, since the pairs involve neutral mesons.

In the experiments of the Belle \cite{Ba07} and LHCb \cite{Aa23}
collaborations, three-dimensional angular distributions in the
decays $D_{s1}(2536)^+\to D^{*+}K^0_S$ and $D_{s1}(2536)^-\to\bar
D^{*0}K^-$ were investigated, respectively, and the ratios of the
amplitudes of the $S$ and $D$ partial waves and their relative
phases were found. Under quite natural theoretical assumptions, the
Belle \cite{Ba07} and LHCb \cite{Aa23} data allow us to obtain the
following estimate for the $S$-wave $D^*K$ scattering length in the
channel with isospin $I=0$ (it is denoted as $a^{(0)}_{D^*K}$):
$|a^{(0) }_{D^*K}|= (0.94\pm0.06)$ fm. Note that this estimate is
close in magnitude to the value $0.76$ fm calculated earlier in the
work \cite{Li11} based on chiral perturbation theory for heavy
mesons.

The paper is organized as follows: the main Section II consists of
three subsections. Subsection II A presents the data from the Belle
and LHCb experiments and provides an estimate of the absolute value
of the $S$-wave amplitude in the $D_{s1}(2536)\to D^*K$ decay. In
Sec. II B, we formulate the assumptions that allow us to obtain an
estimate for $|a^{(0)}_{D^*K}|$ from the Belle and LHCb data, and
present the result itself. In Sec. II C, we discuss the possibility
of explaining the scattering length found by the contribution of the
$D_{s1}(2460)$ resonance. In the final Sec. III, we briefly discuss
the $B_{s1}(5830)\to B^*\bar K$ decay, which is closely related to
$D_{s1}(2536)\to D^*K$.

%------------------------------------------------------------------------------------------------------------------------------------------------
\section{The \boldmath $D^*K$ scattering length from the Belle and LHCb data}

\subsection{The Belle and LHCb data}

Let us briefly recall the data on which our estimate of the
scattering length $a^{(0)}_{ D^*K}$ is based. In the 2008 Belle
experiment \cite{Ba07}, the reaction $e^+e^-\to D_{s1}(2536)^+X$ has
been investigated. In the helicity formalism, the three-dimensional
differential angular distribution in the decay chain $D_{s1}(2536
)^+\to D^{*+}K^0_S$, $D^{*+}\to D^0\pi^+$ was presented in the form
\cite{Ba07}
\begin{eqnarray}\label{Eq1}
\frac{d^3N}{d(\cos\alpha)d\beta d(\cos\gamma)}=\frac{9}{4\pi(1+2
R_\Lambda)}\times\left(\cos^2\gamma\left[\rho_{00}\cos^2\alpha+
\frac{1-\rho_{00}}{2}\sin^2\alpha\right]\right.\nonumber\\
+R_\Lambda\sin^2\gamma\left[\frac{1-\rho_{00}}{2}\sin^2\beta+\cos^2
\beta(\rho_{00} \sin^2\alpha+\frac{1-\rho_{00}}{2}\cos^2\alpha)
\right]\qquad\nonumber\\ \left.+\frac{\sqrt{R_\Lambda}
(1-3\rho_{00})}{4}\sin2\alpha \sin2\gamma\cos\beta\cos\xi\right),
\qquad\qquad\qquad\end{eqnarray} where the angles $\alpha$ and
$\beta$ are measured in the $D^+_{s1}$ rest frame: $\alpha$ is the
angle between the boost direction of the $e^+e^-$ center of mass and
the $K^0_S$ momentum, while $\beta$ is the angle between the plane
formed by these two vectors and the $D^+_{s1}$ decay plane. The
third angle, $\gamma$, is defined in the $D^{*+}$ rest frame between
$\pi^+$ and $K^0_S$. An illustrative description of the kinematics
of the $D_{s1}(2536)^+\to D^{*+}K^0_S\to D^0\pi^+K^0_S$ decay is
given in Fig. 4 in Ref. \cite{Ba07}. Equation (\ref{Eq1}) depends on
three variables: $\rho_{00}$, $R_\Lambda$, and $\xi$ (via
$\cos\xi$). Here $\rho_{00}$ is the diagonal element of the helicity
density matrix of the $D_{s1}(2536)^+$, $\sqrt{R_\Lambda}e^{i\xi}
=A_{1,0}/A_{0,0}=z$, where $A_{1,0}$ and $A_{0,0}$ are the helicity
amplitudes corresponding to the $D^{*+}$ helicities $\pm1$ and 0,
respectively (here we retain the notation adopted in Ref.
\cite{Ba07}). They are related to $S$- and $D$-wave amplitudes in
$D_{s1}(2536)^+$ decay by $A_{1,0}=\frac{1}{\sqrt{3}}(S+\frac{1}
{\sqrt{2}}D)$, $A_{0,0}=\frac{1}{\sqrt{3}}(S-\sqrt{2}D)$. Equation
(\ref{Eq1}) allowed the authors of Ref. \cite{Ba07} to extract
$R_\Lambda$ and $\xi$ (or $z$) and $\rho_{00}$ from the $D_{s1}
(2536)^+$ angular distributions and to obtain $D/S=\sqrt{2}(z-1)
/(1+2z)=\sqrt{\Gamma_D/\Gamma_S}e^{i\eta}$, where $\Gamma_{D,S}$ are
the partial widths of the $D_{s1}(2536)^+$ and $\eta$ is the phase
between the $D$- and $S$-amplitudes. Fitting the three-dimensional
angular distribution to the data gave \cite{Ba07}
\begin{eqnarray}\label{Eq2}
z=A_{1,0}/A_{0,0}=\sqrt{R_\Lambda}e^{i\xi}=\sqrt{3.6\pm0.3\pm0.1}
\exp(\pm i(1.27\pm0.15\pm0.05)). \end{eqnarray} Because the angular
distributions are sensitive only to $\cos\xi$, the phase $\xi$ has a
$\pm\xi+2\pi n$ ambiguity, and $A_{1,0}/A_{0,0 }$ is determined up
to complex conjugation. The ratio of the $D$- and $S$-wave
amplitudes was found to be
\begin{eqnarray}\label{Eq3}
D/S=\sqrt{\Gamma_D/\Gamma_S}e^{i\eta}=(0.63\pm0.07\pm0.02)\exp(\pm
i(0.76\pm0.03\pm0.01)).\end{eqnarray} The absolute value of the
relative phase $\eta$ is close to $\pi/4$, $(43.8\pm1.7\pm0.6
)^\circ$. As emphasized in Ref. \cite{Ba07}, the information on the
relative phase $\xi$ (or $\eta$) can be extracted exclusively from
the whole three-dimensional $d^3N/d(\cos\alpha)d\beta d(\cos\gamma)$
distribution. Indeed, the last interference term in Eq. (\ref{Eq1}),
with $\cos\xi$, vanishes after integration over any angle $\alpha$,
$\beta$, or $\gamma$.

In the 2023 LHCb experiment \cite{Aa23}, the $B^0_{(s)}\to D_{s1}
(2536)^\mp K^\pm$ decays have been investigated. The $D_{s1}(253
6)^-$ meson was reconstructed in the $\bar D^{*0}K^-$ decay channel.
The authors also investigated the full three-dimensional
differential decay rate expressed in terms of the helicity
amplitudes $H_+=H_-$ and $H_0$ corresponding to the $\bar D^{*0}$
helicities $\pm1$ and 0, respectively. The ratio $H_+/H_0$ was
expressed as $ke^{i\phi}$, where $k>0$. These parameters were
determined to be
\begin{eqnarray}\label{Eq4}
k=1.89\pm0.24\pm0.06,\ \ |\phi|=1.81\pm0.20\pm0.11\ \mbox{rad}.
\end{eqnarray}
The amplitude ratio between the $S$- and $D$-partial waves,
$S/D\equiv Ae^{iB}$, was determined to be
\begin{eqnarray}\label{Eq5}
A=1.11\pm0.15\pm0.06,\ \ |B|=0.70\pm0.09\pm0.04\ \mbox{rad}.
\end{eqnarray}
According to LHCb \cite{Aa23}, the fraction of $S$-wave component in
$D_{s1}(2536)^+\to D^{*0}K^+$ is $(55\pm7\pm3)$\%, consistent with
the Belle results from its isospin partner $D_{s1}(2536)^+\to D^{*+}
K^0$, in which the $S$-wave fraction is $(72\pm5\pm1)$\%
\cite{Ba07}. For the $D_{s1}(2536)^+\to D^{*0}K^+$ decay channel,
the threshold of which is 7.344 MeV lower than for
$D_{s1}(2536)^+\to D^{*+}K^0$, the noted decrease in the role of the
S-wave is quite expected. Indeed, if, when passing from the
$D_{s1}(2536)^+\to D^{*+}K^0$ channel to the $D_{s1}(25 36)^+\to
D^{*0}K^+$ channel, the $S$-wave amplitude, $|S|$, remains (in the
first approximation) practically unchanged, and the $D$-wave
amplitude, $|D|$, increases by $q^2_2/
q^2_1=(167\,\mbox{MeV}/149\,\mbox{MeV})^2 =1.256$ times, where $q_1$
and $q_2$ are the momenta of the final $K^0$- and $K^+$-mesons in
the rest frame of $D_{s1}(2536)^+$, respectively, then the $S$-wave
fraction $1/(1+|D|^2/|S|^2)$ will decrease from 72\% in $D_{s1}(2536
)^+\to D^{*+}K^0$ to 61\% in $D_{s1}(2536)^+\to D^{*0}K^+$, which is
consistent with the LHCb data within the errors.

Let us now estimate the absolute values of $|S|$ and $\Gamma_S$
using the Belle \cite{Ba07} and LHCb \cite{Aa23} data for the
partial decay channels $D_{s1}(2536)^+\to D^{*+}K^0$ and
$D_{s1}(2536)^+\to D^{*0} K^+$, respectively, and the natural
assumption about the weak dependence of $|S|$ on the momentum in the
region near the threshold [the resonance $R\equiv D_{s1}(2536)^+$
with $I(J^P) =0(1^+)$, the mass $m_R=(2535.11\pm0.06)$ MeV and width
$\Gamma_R=(0.92\pm 0.0 5)$ MeV \cite{PDG24} is located approximately
30 MeV from the $(D^*K)^+$ threshold]. We write the decay width of
$D_{s1}(2536)^+\to D^{*+ }K^0+ D^{*0}K^+$ as
\begin{eqnarray}\label{Eq6}
\Gamma(D_{s1}(2536)^+\to(D^*K)^+)=\frac{|S|^2}{24\pi m^2_R}\left[q_1
(1+|D/S|^2_{\scriptsize \mbox{Belle}})+ q_2(1+|D/S|^2_{\scriptsize
\mbox{LHCb}})\right].
\end{eqnarray}
Substituting here the central values of $|D/S|^2_{\scriptsize
\mbox{Belle}}$ and $|D/S|^2_{\scriptsize \mbox{LHCb}}$ from Eqs.
(\ref{Eq3}) and (\ref{Eq5}), respectively, $q_1=0.149$ GeV and
$q_2=0.167$ GeV, and also using the central values for $\Gamma_R
=(0.92\pm 0.05)$ MeV \cite{PDG24} and $\mathcal{B}(D_{s1}(2536
)^+\to(D^*K)^+)=(71.8\pm9.6 \pm7.0)\%$ (recently measured for the
first time by the BESIII Collaboration \cite{Ab24}), we obtain
$|S|=0.792$ GeV and $\Gamma_S=0.41$ MeV, and also $\Gamma_D=0.25$
MeV. We will return to these estimates in Sec. II C.

%------------------------------------------------------------------------------------------------------------------------------------------------

\subsection{Estimate of the scattering length \boldmath $a^{(0)}_{D^*K}$}

The idea of estimating $a^{(0)}_{D^*K}$ is very simple. The phases
of the amplitudes of the $S$ and $D$ partial waves in the
$D_{s1}(2536 )^+\to (D^*K)^+$ decays arise due to the nonresonant
(background) $D^*K$ interaction in the final state. Their appearance
is guaranteed by the requirement of unitarity \cite{Wa52,Om58,GW64,
Ba65,LL77,As24}. We assume that near the $D^*K$ threshold the phase
differences of the $S$- and $D$-wave amplitudes measured in
$D_{s1}(2536 )^+\to(D^*K)^+$ decays are almost completely determined
by the phase of the $S$-wave amplitude of the background $D^*K$
scattering. Using the scattering length approximation for this
amplitude and taking into account the contribution of the
$D_{s1}(2536)^+$ resonance, we obtain an estimate for
$a^{(0)}_{D^*K}$. When speaking about the nonresonant (background)
part of $D^*K$ scattering, we mean the representation of partial
wave amplitudes in the form {\it background} $+$ {\it resonance},
widely used for processing experimental data, see, for example
\cite{DB62,Co72,Co73,No80, Co80,Mc67,GM67,LL77,Fl72,Pr73, Bu76,As88,
AS94,AS12,Ro18,As24}. Let us now move on to a more detailed
presentation.

%------------------------------------------------------------------------------------------------------------------------------------------------

Consider the amplitude $\langle\vec{q}_f,\nu'|F^{I=0(J^P=
1^+)}_{D^*K}|\vec{q}_z,\nu \rangle$, describing the reaction
$D^*K\to D^*K$ in a channel with isospin $I=0$ and total angular
momentum and parity $J^P=1^+$. Here $\vec{q}_f$ is the momentum of
the final $D^*$ meson in the reaction center-of-mass system,
$\vec{q}_z$ is the momentum of the initial $D^*$ meson directed
along the $z$ axis in the same system; $|\vec{q}_f|=|\vec{q}_z
|\equiv q\,[\mbox{or }q(\sqrt{s})]=\sqrt{s^2-2s(m^2_{D^*}+ m^2_K
)+(m^2_{D^*}- m^2_K)^2}/(2\sqrt{s})$, $s$ is the square of the
invariant mass of the $D^*K$ pair; $\nu$ and $\nu'$ are the
projections of the spins of the initial and final $D^*$ mesons,
respectively, onto the $z$ quantization axis. Here we look aside
from the effects associated with the differences in the masses of
$K^+$, $K^0$ and $D^{*0}$, $D^{*+}$. Let us expand this amplitude
into partial waves in the $|JMlS\rangle=|1Ml1\rangle$ representation
\cite{GW64,Ne66}:
\begin{eqnarray}\label{Eq7}
\langle\vec{q}_f,\nu'|F^{I=0(J^P=1^+)}_{D^*K}|\vec{q}_z,\nu \rangle
\equiv F_{ \nu',\nu}(s,\theta)e^{i(\nu-\nu')\varphi}=\sum_{l',l}
\sqrt{2l+1} C^{1\nu}_{l0,1\nu}C^{1\nu}_{l'\nu-\nu',1 \nu'} Y_{l'
}^{\nu-\nu'}(\theta,\varphi)f_{l',l}(s),\end{eqnarray} where
$\theta$ is the angle between the momenta $\vec{q}_f$ and
$\vec{q}_z$, $\varphi$ is the azimuthal angle of the momentum
$\vec{q}_f$, $C^{JM}_{lm_l,Sm_s}$ is the Clebsch-Gordan coefficient,
$Y_{l}^{m}(\theta,\varphi)$ is the spherical function, $l$ and $l'$
are the relative orbital momenta in the initial and final states,
respectively; for $J^P=1^+$, the orbital momenta $l$ and $l'$ can
independently take values equal to $J-1=0$ and $J+1=2$ (i.e., states
with different orbital momenta are mixed). With respect to these
states, the quantities $f_{l',l}(s)$ form a symmetric (due to $T$
invariance) $2\times2$ matrix composed of the amplitudes of partial
waves describing the coupled channels \cite{GW64,Ne66,BB52,
St57,Gr75,KS82}. In the normalization we have adopted, the reaction
cross section has the form
\begin{eqnarray}\label{Eq8}
\sigma(s)=\int\frac{4\pi}{3}\sum_{\nu,\nu'}|F_{\nu',\nu}(s,\theta)
|^2 \sin\theta d\theta d\varphi=4\pi\sum_{l',l}|f_{ l',l}(s)|^2.
\end{eqnarray}
Explicit expressions for the amplitudes $F_{\nu',\nu}(s,\theta)$ are
given in the appendix.

Let us represent the matrix $f_{l',l}(s)$ as the sum of the
background, $B_{l',l}(s)$, and resonance, $R_{l',l}(s)$,
contributions
\begin{eqnarray}\label{Eq9} f_{l',l}(s)=B_{l',l}(s)+R_{l',l}(s).
\end{eqnarray} The resonance matrix has the Breit-Wigner form
\begin{eqnarray}\label{Eq10}
R_{l',l}(s)=\frac{1}{2q}\,\frac{V_{l'}V_l}{m_R-\sqrt{s}-i\Gamma_R/2},
\end{eqnarray}
where $V_l$ is the vertex function describing the decay of the
resonance $R\to D^*K$ into an $l$ channel, $|V_0|=\sqrt{\Gamma_S}$
and $|V_2|=\sqrt{ \Gamma_D}$. We also write the amplitude
$B_{l',l}(s)$ as $B_{l',l}(s)=(S^B_{l',l}(s)-\delta_{l',l})/(2iq)$,
where $S^B_{l',l}(s)$ is the symmetric background $S$ matrix. We
will assume that the matrix $S^B_{l',l}(s)$ is unitary. This means
that the background is elastic in the sense that it does not lead to
transitions of $D^*K$ into other particles, but acts only ``inside''
the $D^*K$ channel, mixing the $D^*K$ states with different $l$. Due
to the background interaction, the resonance vertex functions
acquire phases. The unitarity relation for the vertex functions
reads \cite{DB62,No80,Mc67,GM67}:
\begin{eqnarray}\label{Eq11}\mbox{Im}V_l=q\sum_{l'}B_{l,l'}(s)V^*_{l'}
\ \ \ \mbox{or}\ \ \ \sum_{l'}S^B_{l,l'}(s)V^*_{l'}=V_l.
\end{eqnarray} If we require that the original amplitude
$f_{l',l}(s)$ as a whole satisfies unitarity, then together with the
relations in Eq. (\ref{Eq11}), the equality
$\Gamma_R=\Gamma_S+\Gamma_D$ must also be satisfied
\cite{DB62,Co72,Co73,No80,Co80,Mc67,GM67}. We will not require
unitarity for $f_{l',l}(s)$ in order to preserve the possibility of
coupling the $D_{s1}(2536)$ resonance with the decay channels other
than $D^*K$ \cite{PDG24,Ab24}.

As is known, the relations in Eq. (\ref{Eq11}) can be resolved and
the phases of the vertex functions $V_0=e^{i\varphi_0}\sqrt{
\Gamma_S}$ and $V_2=e^{i\varphi_2}\sqrt{\Gamma_D}$ can be expressed
through the parameters of the background and the resonance itself
\cite{DB62, Co72,Co73,No80,Co80,Mc67,GM67} (note that the phases
$\varphi_0$ and $\varphi_2$ involve the signs of the effective
coupling constants, which determine $\sqrt{\Gamma_S}$ and
$\sqrt{\Gamma_D}$). We will demonstrate these relations by the
example of the phase difference of interest to us,
$\varphi_0-\varphi_2$, the modulus of which is known from experience
\cite{Ba07,Aa23}. In the case of two coupled channels, for
parametrization of symmetric unitary matrices of type
$S^B_{l',l}(s)$, there are two conventions --- two ways of
introducing three independent parameters for their description (two
real phases and a mixing parameter) \cite{BB52,St57}; see also Refs.
\cite{GW64,Ne66,MM65,Mc67,GM67,Gr75,Co72,KS82}. In terms of the
eigenphases $\delta_0$, $\delta_2$ and the mixing parameter
$\varepsilon$ \cite{BB52}, the background $S$ matrix $S^B_{l',l}(s)$
is represented as follows:
%------------------------------------------------------------------------------------------------------------------------------------------------
\begin{eqnarray}\label{Eq12} S^B_{l',l}(s)=\left(\begin{array}{cc}
e^{2i\delta_0}\cos^2\varepsilon+e^{2i\delta_2}\sin^2\varepsilon & (e^{2i\delta_0}-e^{2i\delta_2})\cos\varepsilon\sin\varepsilon \\
(e^{2i\delta_0}-e^{2i\delta_2})\cos\varepsilon\sin\varepsilon & e^{2i\delta_0}\sin^2\varepsilon+e^{2i\delta_2}\cos^2\varepsilon \\
\end{array}\right)_{l',l}.
\end{eqnarray}
In practice, in the partial wave analysis of two coupled channels,
an alternative representation of the $S$ matrix is most often used
through the so-called bar-phases and bar-mixing parameter
\cite{St57,KS82}:
%------------------------------------------------------------------------------------------------------------------------------------------------
\begin{eqnarray}\label{Eq13} S^B_{l',l}(s)=\left(\begin{array}{cc}
e^{2i\bar\delta_0}\cos2\bar\varepsilon & ie^{i(\bar\delta_0+\bar\delta_2)}\sin2\bar\varepsilon \\
ie^{i(\bar\delta_0+\bar\delta_2)}\sin2\bar\varepsilon & e^{2i\bar\delta_2}\cos2\bar\varepsilon \\
\end{array}\right)_{l',l}.\end{eqnarray}
Thus, substituting Eq. (\ref{Eq13}) into the second relation in Eq.
(\ref{Eq11}) and calculating the product $V_0V^*_2$, we obtain
\begin{eqnarray}\label{Eq14} \sin(\varphi_0-\varphi_2-\bar\delta_0+\bar\delta_2)=
-\frac{\Gamma_S-\Gamma_D}{2\sqrt{\Gamma_S\Gamma_D}}\tan2\bar\varepsilon.\end{eqnarray}
The connection of $\varphi_0-\varphi_2$ with the eigenphases
$\delta_0$, $\delta_2$ and the mixing parameter $\varepsilon$, see
Eq. (\ref{Eq12}), can be obtained from Eq. (\ref{Eq14}) using the
relations $\bar\delta_0+ \bar\delta_2=\delta_0+\delta_2$, $\sin(
\bar\delta_0-\bar\delta_2)= (\tan2\bar\varepsilon)/(\tan2
\varepsilon)$, and $\sin(\delta_0- \delta_2)=(\sin2\bar\varepsilon)
/(\sin2\varepsilon)$ \cite{MM65}. This results in a very cumbersome
expression. All we tried to demonstrate is the dependence of the
relation between the value $\varphi_0-\varphi_2$ and the parameters
of the background on its specific parametrization. However, the
energy dependence of the matrix elements $S_{l',l}(s)$ as such is
naturally independent of the parametrization. For example, at $q\to
0$ the standard threshold behavior should take place \cite{Ne66}:
\begin{eqnarray}\label{Eq15} B_{l',l}(s)=\frac{S^B_{l',l}
(s)-\delta_{l',l}}{2iq}=\mathcal{O}(q^{l+l'})\end{eqnarray} and,
respectively, it follows from Eqs. (\ref{Eq15}) and (\ref{Eq13})
\begin{eqnarray}\label{Eq16} \bar\delta_0=aq,\ \ \ \bar\delta_2=bq^5,
\ \ \ \sin2\bar\varepsilon=cq^3,\end{eqnarray} where $a$, $b$, and
$c$ are some real constants (the phases $\bar\delta_0$ and
$\bar\delta_2$ are determined with an accuracy of $\pi$). Now
suppose that at $\sqrt{s}=m_R$ (i.e., 30 MeV above the $D^*K$
threshold) the background amplitudes have the usual hierarchy of
$S$- and $D$-wave contributions, in which (in the vicinity of the
threshold) the $S$-wave amplitude dominates. Then from Eqs.
(\ref{Eq14}) and (\ref{Eq16}) at $\sqrt{s}=m_R$, we obtain (up to
$\pi$)
\begin{eqnarray} \label{Eq17}\varphi_0-\varphi_2=\bar\delta_0+
\mathcal{O}(q^3).\end{eqnarray} The absolute values of the
difference $\varphi_0-\varphi_2$ measured by Belle \cite{Ba07} and
LHCb \cite{Aa23} are close to each other; see Eqs. (\ref{Eq3}) and
(\ref{Eq5}). By summing quadratically the statistical and systematic
errors in the data and finding by fitting the average of the
absolute value of the phase difference of the $S$ and $D$ partial
amplitudes for these two experiments, we obtain
$|\varphi_0-\varphi_2|=0.75\pm0.03$ [or $|\varphi_0-\varphi_2|=
(43\pm1.7)^\circ$]. Next, assuming the scattering length
parametrization for the phase $\bar\delta_0$, $\bar\delta_0 =aq$,
and setting $q=0.158$ GeV at $\sqrt{s}=m_R$, we find
\begin{eqnarray} \label{Eq18}
|a|=(0.75\pm0.03)/(0.158\mbox{\ GeV})=(0.94\pm0.04)\ \mbox{fm}.
\end{eqnarray}
The value $q=0.158$ GeV corresponds to the average values of the
$D^*$ and $K$ meson masses in the isotopic multiplets
$m_{D^*}=(m_{D^{*0}}+ m_{D^{*+} })/2=2.00856$ GeV and
$m_{K}=(m_{K^+}+m_{K^0})/2=0.495644$ GeV. An estimate of the
contribution of the $D_{s1}(2536)$ resonance [see Eq. (\ref{Eq10})]
to the scattering length shows that it is about 1.5\% of the
contribution of $|a|$ due to the background. We will take into
account the spread in the absolute value of the scattering length
that appears when taking into account the $D_{s1}(2536)$ resonance
contribution by increasing the uncertainty of our estimate. So,
finally, we obtain the following estimate:
\begin{eqnarray} \label{Eq19} |a^{(0)}_{D^*K}|=(4.75\pm0.19)\
\mbox{GeV}^{-1}=(0.94\pm 0.06)\ \mbox{fm}.\end{eqnarray}

As for the noticeable fraction of the $D$-wave amplitude in the
decay of the resonance $D_{s1}(2536)$ into $D^*K$ \cite{Ba07,Aa23}
(despite its proximity to the $D^*K$ threshold, see Sec. II A), this
\cite{Ba07} and similar phenomena \cite{Av94, Be94} find a natural
explanation within the framework of the heavy quark effective theory
(HQET) \cite{Ba07,IW91,GK91,LW92,Go05}. HQET predictions are as
follows: for an infinitely heavy $c$ quark, the state
$D_{s1}(2536)$, in which the $s$ quark is assumed to have moment
$j=3/2$, should decay into $D^*K$ exclusively in the $D$ wave. Its
partner, the state $D_{s1}(2460)$ \cite{PDG24} with $I(J^P)=0(1^+)$
and $j=1/2$, located below the $D^*K$ threshold, should have
exclusively the $S$ wave coupling to the $D^*K$ channel. Effects
that break the symmetry of heavy quarks lead to mixing of these two
states. Even a small admixture of the $S$ wave in the state
$D_{s1}(2536)$ [it is a narrow one] can be quite noticeable in the
decay $D_{s1}(2536)\to D^*K$ against the background of the $D$ wave,
strongly suppressed by the threshold factor. In the next section, we
discuss the possibility of explaining the found value of the
scattering length, if it is negative, by the contribution of the
$D_{s1}(2460)$ resonance.

%------------------------------------------------------------------------------------------------------------------------------------------------

\subsection{Contribution of the \boldmath $D_{s1}(2460)$ resonance}

Let us try to explain the scattering length $a^{(0)}_{D^*K}$ by the
contribution of the $D_{s1}(2460)$ resonance. We assume that the
$D_{s1}(2460)$ is coupled to the $D^*K$ channel predominantly in the
$S$ wave. Thus, we have some approximate model for the nonresonant
background amplitude $B_{0,0}(s)$ discussed in the previous
subsection. Let us write this amplitude [the amplitude of the
process $D^*K\to D_{s1}(2460)\to D^*K$ near the threshold] in the
Flatt\'{e} form \cite{Fl76,Gu18a} (some other parametrizations will
be discussed elsewhere):
\begin{eqnarray} \label{Eq20} \tilde{f}^B_S(s)=\frac{G^2/2}{m_{\tilde{R}}-
\sqrt{s}-|\tilde{\Gamma}_S(m_{\tilde{R}})|/2-i\tilde{\Gamma}_S(\sqrt{s})/2-i
\tilde{\Gamma}_{\scriptsize\mbox{non-}D^*K}/2}.
\end{eqnarray}
Here, $m_{\tilde{R}}$ is the mass of the $\tilde{R}\equiv D_{s1}
(2460)$ \cite{PDG24}, $G$ represents its coupling to the (closed)
$D^*K$ channel, $\tilde{\Gamma}_S (\sqrt{s})=q(\sqrt{s})G^2$ is its
decay width into $D^*K$, $|\tilde{\Gamma}_S(m_{\tilde{R}})|/2$ is
the subtraction term \cite{Gu18a,ADS80,AS19,AS19a,Aa22}, and
$\tilde{\Gamma}_{\scriptsize\mbox{non-} D^*K}$ is the total width of
the $D_{s1}(2460)$ decay to all open non-$D^*K$ channels
($\tilde{\Gamma}_{\scriptsize\mbox{non-} D^*K}<3.5$ MeV
\cite{PDG24}; further, we neglect this value, which is insignificant
for our estimates). If $\sqrt{s}<(m_{D^*}+m_K)$, then
$\tilde{\Gamma}_S(\sqrt{s})\to i|\tilde{\Gamma}_S(\sqrt{s})|$
\cite{FN1}. Note that the $D_{s1}(2460)$ propagator contains the
finite width correction \cite{ADS80,AS19} and the mass
$m_{\tilde{R}}$ in Eq. (\ref{Eq20}) defines the zero of its real
part. The contribution of
$-|\tilde{\Gamma}_S(m_{\tilde{R}})|/2-i|\tilde{\Gamma}_S(\sqrt{s}
)/2$ both below and above the $D^*K$ threshold turns out to be very
significant. The scattering length due to the subthreshold resonance
(with practically zero width) is negative. Therefore, to estimate
$G^2$ using the found value of $|a^{(0)}_{D^*K} |$, we should set at
$\sqrt{s}=(m_{D^*}+m_K)$, as follows:
\begin{eqnarray} \label{Eq21} \frac{G^2/2}{m_{\tilde{R}}-(m_{D^*}+m_K)
-|q(m_{\tilde{R}})|G^2/2}=-|a^{(0)}_{D^*K}|=-4.75\ \mbox{GeV}^{-1},
\end{eqnarray}
see Eqs. (\ref{Eq19}) and (\ref{Eq20}). Hence it follows that
$G^2=3.6$. Then, for $\sqrt{s}$ equal to the mass of $D_{s1}(2536)$,
$m_R$, the decay width of $D_{s1}(2460)\to D^*K$ turns out to be
very large, $\tilde{\Gamma}_S(m_R)\approx570$ MeV, in comparison to
$\Gamma_S=0.41$ MeV for $D_{s1}(2536)$, given at the end of Sec. II
A. Let us make an ad hoc assumption that $\Gamma_D$ for the
$D_{s1}(2536)$ can be estimated to an order of magnitude from the
relation $\Gamma_D=[q(m_R)/\Lambda]^4\tilde{\Gamma}_S(m_R)$, where
$\Lambda$ is the characteristic energy scale in the decay
$D_{s1}(2536)\to D^*K$. Based on $\Gamma_D=0.25$ MeV (see the end of
Sec. II A), we obtain a quite reasonable value of $\Lambda=1.09$
GeV. Thus, at the physical level of rigor, the picture described
above does not contradict the qualitative expectations based on the
symmetry of heavy quarks for $D_{s1}(2460)$ and $D_{s1}(2536)$
mesons \cite{IW91,GK91,LW92,Go05}. If, as a result of measurements
of femtoscopic correlation functions of pairs $D^{*0}K^+$ and
$D^{*+}K^0$ (or by some other method), it turns out that the
scattering length $a^{(0)}_{D^*K}$ is negative and in magnitude of
about 1 fm, then this can already be considered as some additional
evidence in favor of HQET. For any sign of $a^{(0)}_{D^*K}$, the
theoretical situation with the estimates of competing contributions
of different origins will need to be further clarified.

But even in the case of $a^{(0)}_{D^*K}<0$ with the contribution of
the $D_{s1}(2460)$ resonance, things are not so simple. As can be
seen from Eq. (\ref{Eq20}) at $\tilde{\Gamma}_{\scriptsize
\mbox{non-}D^*K}=0$, the phase of the amplitude $\tilde{f}^B_S(s)$
[we will denote it as $\delta^B_S(s)$] at the $D^*K$ threshold is
equal to $180^\circ$. As $\sqrt{s}$ increases, $\delta^B_S(s)$
decreases, remaining in the second quadrant. At the $D_{s1}(2536)$
resonance point, $\delta^B_S(m^2_R)=145.2^\circ$ at $G^2=3.6$ [note
that with increasing $G^2$ the phase $\delta^B_S(m^2_R)$ cannot
become less than $139.5^\circ$, see Eq. (\ref{Eq20})]. How can one,
at least roughly, reconcile the value
$\delta^B_S(m^2_R)=145.2^\circ$ with one of the values
$\varphi_0-\varphi_2=\pm4 3^\circ$? To do this, it is necessary to
additionally require that the relative sign of the effective
coupling constants in the product of the vertex functions $V_0V_2$
in Eq. (\ref{Eq10}) be negative [see the beginning of the paragraph
before Eq. (\ref{Eq12})]. Then $\delta^B_S(m^2_R)=145.2^\circ$ turns
out to be comparable with the value
$\varphi_0-\varphi_2+180^\circ=(-43+ 180)^\circ=137^\circ$.

We note that the above mentioned possibility of obtaining
information about the constant $G^2$, responsible for the coupling
of the $D_{s1}(2460)$ resonance to the $D^*K$ decay channel, based
on the data from femtoscopic experiments measuring the $D^*K$
scattering length, seems to us to be rather unique one.

%------------------------------------------------------------------------------------------------------------------------------------------------

\section{Conclusion}

%------------------------------------------------------------------------------------------------------------------------------------------------

We used information on the relative phases of the $S$- and $D$-wave
amplitudes in the $D_{s1}(2536)^+\to D^{*+}K^0_S$ and $D_{s1}(253
6)^-\to\bar D^{*0}K^-$ decays obtained in the Belle \cite{Ba07} and
LHCb \cite{Aa23} experiments to estimate the absolute value of the
$S$-wave $D^*K$ scattering length in the isospin-0 channel,
$|a^{(0)}_{D^*K}|=(0.94\pm 0.06)\ \mbox{fm}$. If $a^{(0)}_{D^*K}$ is
negative, then its value can, in principle, be explained by the
contribution of the $D_{s1}(2460)$ resonance. Certainly, the
experimental measurement of $a^{(0)}_{D^*K}$ and further studies of
the properties of the $D_{s1}(2460)$ and $D_{s1}(2536)$ mesons are
tasks at the leading edge of charm physics.

In conclusion, let us briefly dwell on the decays $B_{s1}(5830)\to
B^{*+}K^-$ and $B_{s1}(5830)\to B^{*0}\bar K^0$
\cite{PDG24,Wa12,Si18,Liu24}, which are closely related to the
decays $D_{s1}(2536)\to D^*K$. The $B_{s1}(5830)$ resonance is
located approximately 10 and 5.5 MeV from the $B^{*+}K^-$ and
$B^{*0} \bar K^0$ thresholds, respectively. It would be very
interesting to find out whether the $D$-wave contribution is still
noticeable in these decays. In general, a detailed study of the
angular distributions in the decay chains $B_{s1}(5830) \to
B^{*+}K^-\to B^+\gamma K^-$ and $B_{s1}(5830)\to B^{*0}\bar K^0\to
B^0\gamma\bar K^0$ (the corresponding formulas are presented in Ref.
\cite{Aa23}) would make it possible to obtain data on the relative
magnitudes and phases of the amplitudes of the $S$ and $D$ partial
waves. This would greatly facilitate theoretical calculations of the
above amplitudes. It would also be of interest to refine the result
obtained by the CMS Collaboration \cite{Si18} for the ratio
$R^{0\pm}_1=\mathcal{B}(B_{s1}(5830)\to B^{*0}\bar K^0_S)/
\mathcal{B}(B_{s1}(5830)\to B^{*+}K^-)=0.49 \pm0.14$. The point is
that for the $S$-wave contribution, under the assumption of isotopic
invariance, this ratio is approximately equal to $0.365$ due to
threshold factors, while the $D$-wave contribution can only reduce
it. In Ref. \cite{Wa12}, a value of 0.23 was predicted for
$R^{0\pm}_1$, but not due to the contribution of the $D$ wave. The
negative sign of the $I=0$ $S$-wave $B^*K$ scattering length would
serve as an indirect indication of the existence of a state with
quantum numbers $I(J^P)=0(1^+)$, $j=1/2$ (a partner of the $B_{s1}
(5830)$ with $j=3/2$) located below the $B^*K$ threshold, which has
not yet been discovered.

%------------------------------------------------------------------------------------------------------------------------------------------------

\vspace*{0.3cm}

\begin{center} {\bf ACKNOWLEDGMENTS} \end{center}

We are grateful to A. E. Bondar for discussions of the decays
$D_{s1}(2460)\to D_s\pi\pi$ and $D_{s1}(2536)\to D_s\pi\pi$ which
inspired the present work. The work was carried out within the
framework of the state contract of the Sobolev Institute of
Mathematics, Project No. FWNF-2022-0021.

%------------------------------------------------------------------------------------------------------------------------------------------------

\vspace*{0.3cm}

\begin{center} {\bf DATA AVAILABILITY} \end{center}

The data that support the findings of this. article. are openly
 available [18, 19].

%------------------------------------------------------------------------------------------------------------------------------------------------

\vspace*{0.3cm}

\begin{center} {\bf APPENDIX:\, THE AMPLITUDES {\boldmath
$F_{\nu',\nu}(s,\theta)$}} \end{center}

\setcounter{equation}{0}
\renewcommand{\theequation}{A\arabic{equation}}

%------------------------------------------------------------------------------------------------------------------------------------------------

The amplitudes $F_{\nu',\nu}(s,\theta)$ defined in Eq. (\ref{Eq7})
have the form
\begin{eqnarray}\label{A1}
F_{1,1}(s,\theta)=F_{-1,-1}(s,\theta)=\frac{1}{\sqrt{4\pi}}\left[f_{0,0}(s)+\frac{1}{\sqrt{2}}f_{0,2}(s)\right]+
\frac{1}{2\sqrt{5}}Y_{2}^{0}(\theta)[f_{2,2}(s)+\sqrt{2}f_{2,0}(s)],
\end{eqnarray}
\begin{eqnarray}\label{A2}
F_{0,0}(s,\theta)=\frac{1}{\sqrt{4\pi}}[f_{0,0}(s)-\sqrt{2}f_{0,2}(s)]+\frac{2}{\sqrt{5}}Y_{2}^{0}(\theta)
\left[f_{2,2}(s)-\frac{1}{\sqrt{2}}f_{2,0}(s)\right],
\end{eqnarray}
\begin{eqnarray}\label{A3}
F_{1,-1}(s,\theta)=F_{-1,1}(s,\theta)=\sqrt{\frac{3}{10}}Y_{2}^{2}(\theta,0)
[f_{2,2}(s)+\sqrt{2}f_{2,0}(s)],
\end{eqnarray}
\begin{eqnarray}\label{A4}
F_{0,1}(s,\theta)=-F_{0,-1}(s,\theta)=-\frac{1}{2}\sqrt{\frac{3}{5}}Y_{2}^{1}(\theta,0)
[f_{2,2}(s)+\sqrt{2}f_{2,0}(s)],
\end{eqnarray}
\begin{eqnarray}\label{A5}
F_{1,0}(s,\theta)=-F_{-1,0}(s,\theta)=\sqrt{\frac{3}{5}}Y_{2}^{1}(\theta,0)
\left[f_{2,2}(s)-\frac{1}{\sqrt{2}}f_{2,0}(s)\right],
\end{eqnarray}
There is a supplementary relation between the amplitudes $F_{\nu'
\nu}(s,\theta)$: $F_{1,1}-F_{1,-1}-F_{0,0}=\sqrt{2}\cot\theta (F_{0
,1}+F_{1,0})$, which is the $T$-invariance consequence \cite{GW64}.
In terms of the amplitudes $f_{l',l}(s)$, it is reduced to the
equality $f_{0,2}(s)=f_{2,0}(s)$.

%------------------------------------------------------------------------------------------------------------------------------------------------

%------------------------------------------------------------------------------------------------------------------------------------------------


\begin{thebibliography}{99}
\bibitem{FN1} The expressions for the polarization operator of a resonance
$\tilde{R}$, which correspond to the contribution of the $ab$
intermediate state ($\tilde{R}\to ab\to\tilde{R}$), in different $s$
regions [i.e., for $s>m^2_+$, $m^2_-<s<m^2_+$, and $s<m^2_-$, where
$m^2_\pm=(m_a\pm m_b)^2$] are related to each other by analytic
continuation \cite{ADS80,AS19,AS19a,Aa22}. Here we mean the
polarization operator with the real part subtracted at
$s=m^2_{\tilde{R}}$ and written using the usual relativistic
formulas \cite{ADS80,AS19,AS19a,Aa22}. A comparison of the
relativistic formulas for the polarization operator with its
nonrelativistic approximation of the form $G^2[-|q(m_{\tilde{R}}
)|-iq(\sqrt{s})]/2$ used by us in Eq. (20) (see also, for example,
Eq. (16) in Ref. \cite{Gu18a}) for $s$ in the vicinity of the
threshold $m^2_+=(m_{D^*}+m_K)^2$ shows that such an approximation
is completely justified at least in the interval $(m_+-40\,
\mbox{MeV}) <\sqrt{s}<(m_++40\,\mbox{MeV})$; in so doing,
$q(\sqrt{s})=\sqrt{ (s-m^2_+)(s-m^2_-)}/(2\sqrt{s})\to i|q(\sqrt{s}
)|$ if $m^2_-<s <m^2_+$. In fact, Eq. (20) is used by us in
precisely this interval of $\sqrt{s}$, where it has good unitary and
analytical properties \cite{Fl76,Gu18a,ADS80,AS19,AS19a,Aa22}. Note
that the threshold $m_+=m_{D^*}+ m_K$ and the pseudothreshold
$m_-=m_{D^*}-m_K$ are separated by $2m_K\approx1$ GeV and, of
course, the nonrelativistic approximation is already inapplicable on
such intervals.
\end{thebibliography}
\end{document}